\documentclass[twocolumn,english,aps,prl,showpacs]{revtex4}
\usepackage{graphicx}
\usepackage{babel}
 

\begin{document}

\title{Quantum Hall fractions in ultracold fermionic vapors}

\author{N.~Regnault}
\email{Nicolas.Regnault@lpa.ens.fr}

\author{Th.~Jolicoeur}
\email{Thierry.Jolicoeur@lpa.ens.fr}

\affiliation{Laboratoire Pierre Aigrain, D\'epartement de
Physique, 24, rue Lhomond, 75005 Paris, France}


\begin{abstract}
We study the quantum Hall states that appear in the dilute limit
of rotating ultracold fermionic gases when a single hyperfine
species is present. We show that the $p$-wave scattering
translates into a pure hard-core interaction in the lowest Landau
level. The Laughlin wavefunction is then the exact ground state at
filling fraction $\nu$=1/3. We give estimates of some of the gaps
of the incompressible liquids for $\nu = p/(2p\pm 1)$. We estimate
the mass of the composite fermions at $\nu =1/2$. The width of the
quantum Hall plateaus is discussed by considering the equation of
state of the system.
\end{abstract}

\pacs{03.75.Ss, 05.30.Fk, 73.43.-f, 73.43.Cd}

\maketitle



The preparation and manipulation of ultracold atomic gases has led
to many interesting developments in the study of quantum fluids
undergoing fast rotation~\cite{Matthews99,Haljan01}. The
Bose-Einstein condensates can be set in various rotation regimes
with the characteristic response of a superfluid. The condensate
does not acquire angular momentum below some velocity threshold.
Then there is nucleation of one vortex and with increasing
velocity more and more vortices are created. They have been
observed forming the Abrikosov triangular
lattice~\cite{Madison00,Abo01}. When the rotation frequency
reaches the trapping frequency in the radial plane, it has been
predicted that quantum Hall fractional states should become ground
states of the system~\cite{Wilkin98,Cooper99,Wilkin00,Cooper01} if
the gas enters a two-dimensional regime. Trapped Fermi gases may
also exhibit superfluidity if they undergo BCS pairing
condensation. If the pairing strength is varied, it may be
possible to observe the crossover from molecular condensation at
strong coupling to BCS phase transition at weak coupling. A
possible signature of the superfluid paired phase of fermions is
the peculiar response to stirring, leading again to vortex
formation. In the fast rotation limit, it is thus natural to ask
if there is formation of fractional quantum Hall states as in the
Bose case and what are their properties.


In this Letter we investigate the fractional quantum Hall effect
(FQHE) appearing in atomic vapor made of a single hyperfine
species of fermions. We show that the $p$-wave scattering between
fermions can lead to the formation of the Jain principal sequence
of FQHE fractions $\nu =p/(2p\pm 1)$, in addition to the
celebrated Laughlin wavefunction at $\nu =1/3$, as well as a Fermi
sea of composite fermions for half-filling of the lowest Landau
level (LLL). We give estimates of the gaps for the incompressible
fluids governed by the $p$-wave scattering length and of the mass of
the composite fermions. The equation of state of the system seen
as the angular momentum of the ground state as a function of the
rotation frequency displays plateaus corresponding to the FQHE
fluids. Their widths can be estimated by taking into account the
nucleation of quasiparticles.

We consider a gas of fermionic atoms and suppose that they are set
in rotation for example by a stirring external potential~\cite{Rosenbusch02}
that can
be applied for some time to transfer angular momentum to the gas
and then is removed. We are then left with a rotating cloud and we
assume that it attains thermal equilibrium in the rotating frame.
If $\mathcal{H}$ stands for the Hamiltonian in the laboratory
frame then it becomes $\mathcal{H}_R =\mathcal{H}-\omega L_z$ in
the rotating frame where $\omega$ is the rotation frequency and
$L_z$ the angular momentum along the rotation axis.
The Hamiltonian describing N particles of
mass $m$ in this frame can be written as~:
\begin{eqnarray}\label{Ham1}
  \mathcal{H}_R &=& \sum^{N}_{i=1}\frac{1}{2m}(\mathbf{p}_{i}-m\omega
    \mathbf{\hat{z}}\times \mathbf{r}_{i})^{2}
    +\frac{1}{2}m\omega_{z}^{2}z_{i}^{2} \\
   &+&\frac{1}{2}m(\omega_{0}^{2}-\omega^{2})(x_{i}^{2}+y_{i}^{2})
+\sum_{i<j}^{N}V(\mathbf{r}_{i}-\mathbf{r}_{j}),
\nonumber
\end{eqnarray}
where the $xy$ trap frequency is $\omega_{0}$, the axial frequency
is $\omega_{z}$ and the angular velocity vector is
$\omega\mathbf{\hat{z}}$. For $\omega$ close to $\omega_0$, the
physics is that of charge-$e$ particles in a magnetic field
$\mathbf{B}=(2m\omega/e)\mathbf{\hat{z}}$, corresponding to a
magnetic length $\ell=\sqrt{\hbar/(2m\omega)}$. We assume the
existence of a two-dimensional (2D) regime in which the
wavefunction along the $z$-axis is the ground state of the
$z$-axis harmonic potential.

If we consider a single hyperfine species of fermions, then the
$s$-wave scattering is forbidden by the Pauli principle. The next
allowed partial wave, the $p$-wave, leads to much weaker
interactions~\cite{DeMarco99} and this leads to difficulties when cooling fermionic
vapors. They can be evaded for example by sympathetic
cooling~\cite{Myatt97} with a different atom. However it is also
feasible to use a scattering resonance, such as a Feshbach
resonance, to dramatically enhance $p$-wave scattering. This has
been demonstrated with $^{40}$K atoms~\cite{Regal03}. The
scattering even reaches values comparable to $s$-wave scattering.
We will see that this means that FQHE gapped states will have
characteristic energies in the same range as for similar bosonic
states. At small wavevector, i.e. in the low-energy limit,
the $p$-wave phase shift of the two-body scattering problem behaves as~:
\begin{equation}\label{phaseshift}
\delta_1 (k)\,\sim \,\frac{1}{3}\, k^3\, a_1^3 ,
\end{equation}
where $a_1$ defines the $p$-wave scattering length. As a consequence
the scattering amplitude is no longer isotropic~:
\begin{equation}\label{amplitude}
    f_1 (\theta)\,\sim\, a_1^3\, k^2 \,\cos\theta ,
\end{equation}
where $\theta$ is the angle between ingoing and outgoing
wavevectors. We now use an effective potential which mimics the
behavior in Eq.(\ref{amplitude}) when treated in the Born
approximation. It is given by~:
\begin{equation}\label{pseudo}
    \hat{\mathcal{U}}_p = \frac{12\pi a_1^3}{m}\, \hat{\mathbf{p}}\,\,
    \delta^{(3)}(\mathbf{r})\,\,
    \hat{\mathbf{p}},
\end{equation}
where the quantities $\mathbf{r}=\mathbf{r}_1-\mathbf{r}_2$ and
$\mathbf{p}=\frac{1}{2}(\mathbf{p}_1-\mathbf{p}_2)$ pertain to the
relative particle. The use of such a potential is enough for our
purpose since we will study the dynamics of the LLL only~: the
massive degeneracy is lifted at first-order in the potential
(higher orders involve Landau level mixing).

We now turn to the quantum Hall regime for fermions when $\omega
=\omega_0$ in Eq.(\ref{Ham1}). Assuming a 2D regime with
$\ell_z=\sqrt{\hbar/m\omega_z}$ the confinement length along $z$,
the interaction potential can be written as $g_f\,\, \ell^4\,
\hat{\mathbf{p}}\,\,\delta^{(2)}(\mathbf{r})\,\,\hat{\mathbf{p}}$
where the vectors $\mathbf{r}$ and $\mathbf{p}$ are now 2D with~:
\begin{equation}\label{coupling}
    g_f= \sqrt{\frac{2}{\pi}}\frac{\hbar^2}{m}\frac{a_1^3}{\ell_z\ell^4}.
\end{equation}
The coupling constant $g_f$ sets the scale of the FQHE phenomenon.

The interaction Hamiltonian in the LLL can be written as~:
\begin{equation}\label{Proj}
    \mathcal{H}_{LLL}=\sum_{m}\sum_{i,j}V_m \,\, \hat{P}_m (i,j),
\end{equation}
where $m$ is the relative angular momentum (RAM), hence the sum
runs over all positive odd integers for fermions, and $\hat{P}_m
(i,j)$ is the projection operator for particles $i$ and $j$ onto
RAM $m$. The coefficients $V_m$ fully characterize the interaction
problem in the LLL. They are called pseudopotentials after
Haldane~\cite{Haldane83,HaldaneBook85}. With the pure $p$-wave
interaction Eq.(\ref{pseudo}), only $m=1$ scattering is allowed
and the interaction Hamiltonian reduces immediately to the pure
hard-core model for which $V_1\neq0$ and all the other
pseudopotentials are zero. The hard-core
model~\cite{Haldane83,HaldaneBook85,Trugman85,MacDonald94} is
known to constitute an excellent blueprint of the FQHE. Notably
the celebrated Laughlin wavefunction~\cite{Laughlin83}~:
\begin{equation}\label{LJ}
    \Psi =\prod_{i<j}(z_i-z_j)^3\,\, \mathrm{e}^{-\sum_i |z_i|^2/4\ell^2}
\end{equation}
is an exact zero-energy ground state of the hard-core model. In
fact this is the most spatially compact zero-energy state for the
hard-core model. It has a well-defined angular momentum $L_z
=3N(N-1)/2$. It corresponds to filling 1/3 of the LLL and thus is
the exact ground state for ultracold fermions. This is analogous
to the Bose problem~\cite{Wilkin00} where $s$-wave scattering
leads to the Laughlin ground state (with exponent 2) at filling
1/2. It is also known that the hard-core model exhibits the
prominent Jain sequence~\cite{Jain89} of incompressible fluids for
fillings of the form $\nu =p/(2p\pm 1)$. We have estimated the
gaps at some of these fractions by performing exact
diagonalizations in the spherical
geometry~\cite{Haldane85,Fano86,Morf02}. A sphere of radius $R$ is
threaded by a flux $4\pi R^2 B$ which is an integral multiple (2S)
of the flux quantum by Dirac quantization condition.
Incompressible fluids appear for special matching of the number of
particle vs. 2S. We obtain the low-lying energy levels for a small
number of particles and then perform finite-size scaling to obtain
estimations of the thermodynamic limit.

\begin{figure}[!htbp]
\begin{center}\includegraphics[  width=3.25in,
 keepaspectratio]{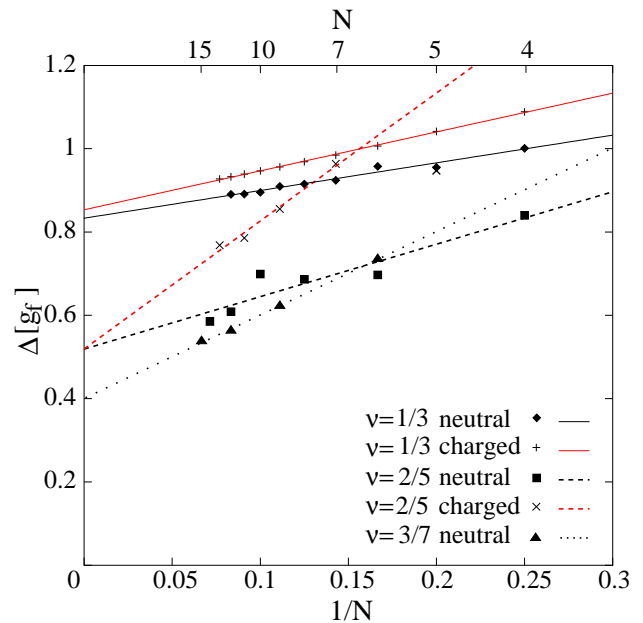}\end{center}
\caption{Energy gaps for neutral and charged excitations at $\nu =1/3$ , $\nu
=2/5$, and neutral gaps for $\nu =3/7$. The lines are our best fits and
energies are in units of $g_f$.}
\end{figure}

For the most stable fluid at $\nu=1/3$ the low-lying neutral
excited states are dominated by a well-defined collective density
mode and we have obtained the corresponding gap by studying up to
N=13 fermions. The gap can also be evaluated from the charged
excitations, i.e. the Laughlin quasiparticles. For the $p$-wave
problem, there is the peculiarity that the quasiholes are gapless
at $\nu =1/3$. It is thus enough to study the quasielectrons that
can be nucleated by removing one flux quantum from the reference
$\nu =1/3$ situation. We find that both estimates scale nicely to
a common value $\simeq 0.8\, g_f$~: see fig.(1). For the fraction
$\nu =2/5$ the gap is $\simeq 0.5\, g_f$ obtained by the study of
systems for N=4...12 fermions. In this case there are the two
types of charged excitations, quasiholes and quasielectrons, each
having a nonzero gap and we find that the sum of these gaps
converges to a value compatible with the neutral gap. There are
fewer available values of the number of particles as we go down
the hierarchy and for the next fraction $\nu =3/7$ we estimate the
gap to be  $\simeq 0.4\, g_f$ and for 4/9 it is more difficult to
give a reliable estimate, the gap is  smaller and of the order of
$\simeq 0.3\, g_f$. For values of the parameters $\ell_z$, $\ell$
and $a_1$ typical of present experiments~\cite{Madison00,Regal03}
the gaps may be of the order of the nanoKelvin.

For fillings less than 1/3, the fermion system does no longer
display incompressibility because there are proliferating
zero-energy states when we increase the angular momentum. Some of
these are edge excitations of the droplet and for larger angular
momentum they are quasiholes. Due to particle-hole symmetry, these
very same modes fill any gap in the region $1\geq\nu\geq 2/3$.
\begin{figure}[!htbp]
\begin{center}\includegraphics[  width=3.0in,
 keepaspectratio]{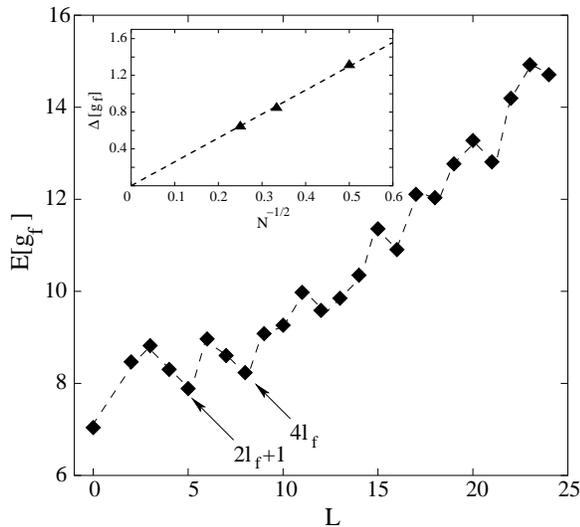}\end{center}
\caption{Energy spectrum for N=9 fermions in the spherical geometry at filling $\nu =1/2$.
The CFs feel zero flux and can be interpreted as forming a closed shell.
The inset shows the scaling of the gap of the closed shell states N=4,9,16.
Energies are in units of $g$ and the horizontal axis
is total angular momentum.}
\end{figure}
The Jain sequence has an appealing interpretation in terms of
composite fermions (CF). These entities are naively a fermion
bound to an even number of flux quanta of a fictitious field. The
total field acting upon the CFs is then the sum of the external
field and the fictitious field~: when treated in a mean-field
manner this explains the FQHE of electrons as the integer quantum
Hall effect of CFs. The Jain sequence has an accumulation point at
$\nu =1/2$~: at this filling the CF experience zero net flux and
form a Fermi liquid like ground state~\cite{Halperin93,Lopez91}.
These CFs have remaining interactions and also an effective mass
$m^{*}$ which is entirely due to interactions. To estimate this
mass, we use a special matching of the flux $2S=2(N-1)$ giving
zero net field on the sphere~\cite{RezayiRead,Morf95} for the CFs.
In the Coulomb case, it has been shown that many features of the
spectrum can be successfully interpreted by reasoning with free
CFs, eventually supplemented by second Hund's rule. There are
closed shell configurations when the number of fermions is a
square~: N=($\ell$+1)$^2$ and $\ell$ is thus the total angular
momentum of the highest occupied orbital. It is the equivalent of
the Fermi momentum on the sphere and thus we call it $\ell_F$. We
expect that the closed shell sequence
N=($\ell_F$+1)$^2$=4,9,16,\dots display ground states with zero
total angular momentum and should have good scaling properties
towards the thermodynamic limit, as is the case for Coulomb
interactions~\cite{Morf95,Onoda00}. From a closed-shell
configuration one can form particle-hole (ph) excitations~: the
lowest-lying such excitations is obtained by promoting a fermion
from the shell with momentum $\ell_F$ to the empty shell at
momentum $\ell_F$+1, leading to a branch extending from L=1 up to
2$\ell_F$+1. Above this branch we should find
two-particle-two-hole states extending up to 4$\ell_F$ and so on.
This is exactly what we find for the hard-core model. The
low-lying levels of N=9 fermions are displayed in fig.~2. Above
the singlet ground state we clearly identify the two branches
predicted by the free CF model.

We can obtain an estimate of the CF mass at $\nu =1/2$ by using
the free CF model~\cite{Morf95,Onoda00}. A free CF on the sphere
has an energy given by $E= l(l+1)/(2m^* R^2)$ where $l$ is the
angular momentum. As a consequence, the gap of the one ph branch
is $\Delta =(\ell_F+1)/(m^* R^2)$. If we fix the density $\rho$
the scaling law of the gap becomes $\Delta_N
=4\pi\rho\sqrt{N}/(m^*(N-1))$. In the spherical geometry we have
to take into account the fact that there is a nontrivial shift
between the flux $2S$ and the thermodynamic limit value $N/\nu$.
For finite numbers of particles the density is not exactly equal
to the thermodynamic limit value. Better scaling
properties~\cite{Morf02} are obtained by rescaling the magnetic
length by a factor $\sqrt{N/(2S\nu)}$ (going to unity for
$N\rightarrow\infty$). We thus find $\Delta_N
=4\pi\rho/(m^*\sqrt{N})$. This scaling is obeyed for the sizes
N=4,9,16 (see inset of fig.(2)) and this leads to an estimate of
the effective mass~:
\begin{equation}\label{mass}
    m^* \simeq 0.5\,\, m\, \frac{\ell^2 \ell_z}{a_1^3}.
\end{equation}
When the number of fermions lie between closed shell values we
have checked that the ground state angular momentum is given by
second Hund's rule (maximum L), as is the the case for Coulomb
interactions~\cite{RezayiRead}.


Finally we discuss the width of the Hall plateaus in the case of
trapped atomic vapors. Contrary to the 2D electron systems in
semiconductor devices there is no source of disorder to pin the
quasiparticles that are nucleated when we deviate from the
fine-tuning of a quantum Hall fraction. The role of disorder is
thus played by the finite number of particles of the system and
the quantum Hall plateaus are expected to be of vanishingly small
width in the thermodynamic limit. We can give precise estimates by
considering the equation of state of the rotating system, i.e. the
value of the ground state angular momentum as a function of the
rotation frequency $\langle L_z\rangle (\omega)$. At the critical
frequency, the Hamiltonian is rewritten as a magnetic field
problem (see Eq.(\ref{Ham1})). If the frequency is slightly less,
$\omega_0 -\delta\omega $, then we have the small field
$-\delta\omega L_z$ acting upon the purely magnetic problem. It
leads to a trivial shift of the energies of the FQHE problem that
will change the ground state when increased. This is seen from the
typical spectrum displayed in fig.~(3) in the condition of the
critical rotation. The Laughlin state at $L_z^{Laughlin}
=3N(N-1)/2$ is the ground state and the first excited state is the
quasielectron at $L_z^{Laughlin}-N$ with a nonzero gap
$\Delta_{qe}$. When adding a $-\delta\omega L_z$ shift the
Laughlin will remain the ground state till the quasielectron
energy becomes lower for a critical value equal to~:
\begin{equation}\label{plateau}
\delta\omega_{c}=\frac{\Delta_{qe}}{N\hbar}\simeq \frac{10^2\,\,
Hz }{N}\,\,\times (\frac{\Delta_{qe}}{1\,\mathrm{nK}}).
\end{equation}
If we increase $ \delta\omega$ beyond this value, quasielectrons are nucleated
forming a fluid that will condense into a new FQHE fluid. We expect the result above
in Eq.(\ref{plateau}) to be generic. This picture is essentially dual
to the nucleation of vortices at small rotation frequency~\cite{Madison00}.

\begin{figure}[!htbp]
\begin{center}\includegraphics[  width=3.0in,
 keepaspectratio]{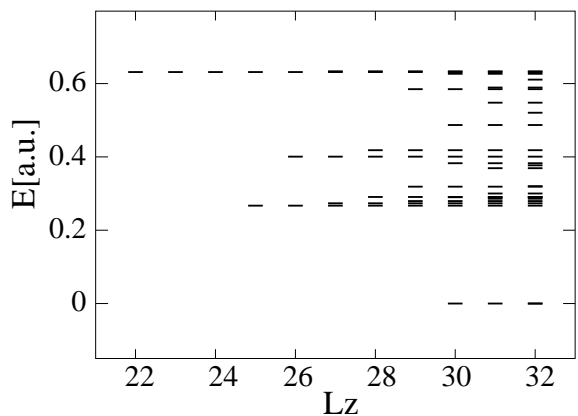}\end{center}
\caption{Energy spectrum for N=5 fermions in the disk geometry
as a function of the angular momentum. The Laughlin state is the unique
zero-energy state at $L_z =30$. The lowest-energy state at $L_z =25$
is the quasielectron with a finite gap.}
\end{figure}

We have shown the appearance of the Jain principal sequence of
quantum Hall fractions in ultracold rotating fermionic vapors. The
composite fermion picture gives a successful account of the
observed fractions as well as their collective mode excitations.
The gaps we estimate from exact diagonalizations are
 of the order of $\hbar^2 a_1^3/m\ell_z\ell^4$.
At half-filling of the lowest-Landau level, there is a Fermi liquid-like
state of composite fermions and their effective mass $m^*/m$ is
controlled by $\ell^2 \ell_z /(a_1^3)$.


\begin{acknowledgments}
We thank  Yvan Castin and Jean Dalibard for numerous discussions.
We thank also IDRIS-CNRS for a computer time allocation.
\end{acknowledgments}


\end{document}